\documentstyle[preprint,eqsecnum,aps,prb]{revtex}

\begin{document}
\draft
\title{Influence of high-energy electron irradiation \\ on the transport 
properties of \\ La$_{1-x}$Ca$_{x}$MnO$_{3}$ films ($x \approx 1/3$).
}
\author{B. I.  Belevtsev$^{*}$ and V. B.  Krasovitsky}
\address{B. Verkin Institute for Low Temperature 
Physics \& Engineering, Kharkov, 310164, Ukraine
}
\author{V. V.  Bobkov}
\address{Kharkov State University, Kharkov, 310077, Ukraine
}
\author{D. G. Naugle$^{\dag}$, K. D. D. Rathnayaka and A. Parasiris}
\address{Department of Physics, Texas A\&M University, 
College Station, TX 77843-4242, USA
}
\maketitle
\begin{abstract}
The effect of crystal lattice disorder on the conductivity and colossal 
magnetoresistance in La$_{1-x}$Ca$_{x}$MnO$_{3}$~($x \approx 0.33$) films has been
examined. The lattice defects are introduced by irradiating the film with high-energy
($\simeq 6$~MeV) electrons with a maximal fluence of about $2\times 10^{17}$~cm$^{-2}$.
This comparatively low dose of irradiation produces rather small radiation damage in
the films. The number  of displacements per atom (dpa) in the irradiated sample is
about $10^{-5}$. Nevertheless, this results in an appreciable increase in the film
resistivity. The percentage of the resistivity increase in the ferromagnetic metallic
state (below the Curie temperature $T_{c}$) was much greater than that observed in the
insulating state (above $T_{c}$). At the same time irradiation has much less effect
on $T_{c}$ or on the magnitude of the colossal magnetoresistance. A possible
explanation of such behavior is proposed.   

\end{abstract}
\section{Introduction}
\par
In recent years considerable attention has been focussed on 
the structural, magnetic and electron transport properties of perovskite 
oxides of the type 
R$_{1-x}$A$_{x}$MnO$_{3}$ (where $R$ is a rare-earth element, 
$A$ a divalent alkaline-earth element). This interest was caused 
by observation of an extremely large negative magnetoresistance  in 
these compounds \cite{helm93,jin94}, which was called colossal 
magnetoresistance (CMR).  Along with fundamental importance for 
condensed matter physics, this phenomenon also offers applications 
in advanced technology. Therefore the problem of CMR 
continues to be topical. 
\par
The doped manganites undergo a phase
transition with decreasing temperature from a paramagnetic insulating
state into a highly conducting ferromagnetic phase. It can be said that 
this insulator-metal transition occurs approximately simultaneously 
with a paramagnetic-ferromagnetic transition (at least in good quality
crystals). The external magnetic field shifts the transition temperature
$T_{c}$ (which is usually near room temperature in well ordered samples 
with $x \approx 0.33)$ to higher temperature producing the CMR 
(see reviews in Refs.\ \onlinecite{nagaev,coey,ram,khom}).
\par
The most pronounced CMR effect was found in 
La$_{1-x}$Ca$_{x}$MnO$_{3}$ films with $x \simeq 1/3$.
The undoped compounds from this series (LaMnO$_{3}$ and CaMnO$_{3}$) 
are antiferromagnetic insulators. In the intermediate range of doping 
($0.2 < x < 0.4$) La$_{1-x}$Ca$_{x}$MnO$_{3}$ is a ferromagnetic 
conductor at low temperature. The ferromagnetic state
is believed to be due to the appearance of Mn$^{4+}$ ions with 
substitution of La$^{3+}$ by a divalent cation. It can be assumed 
that ferromagnetism 
results from the strong ferromagnetic exchange betweeen 
Mn$^{3+}$ and Mn$^{4+}$. The appearance of such an interaction can be 
qualitatively explained within the double-exchange (DE) model 
\cite{zener,and55,gennes}. This model, however, cannot alone explain 
either the huge drop in resistance at the transition, or the real nature 
of the insulating state at $T > T_{c}$ and, therefore, the conductivity 
mechanism in this state. Thus, additional physical processes have been 
invoked to explain the insulating state and insulator-metal transition.
Among them are lattice (polaron) effects \cite{millis} and the possibility 
of phase separation into charge-carrier-poor and charge-carrier-rich 
regions\cite{nagaev,khom,gor}.
\par
The conductivity of La$_{1-x}$Ca$_{x}$MnO$_{3}$ with $x < 0.5$ is 
determined by holes which appear as the result of replacement of 
trivalent La by divalent atoms. The DE model is based on the assumption 
that the holes in doped manganites correspond to Mn$^{4+}$ ions 
arising among the regular Mn$^{3+}$ ions due to doping. However another 
point of view exists\cite{nagaev,khom,alex} that the holes go on oxygen 
sites. The experimental data on this point are contradictory. There is 
experimental evidence (see Ref.\ \onlinecite{croft} and references therein) 
that holes doped into LaMnO$_{3}$ are mainly of Mn $d$ character. 
On the other hand experimental studies described in 
Refs.\ \onlinecite{saitoh,ju} give evidence that the charge carriers 
responsible for conduction in doped manganites have
significant oxygen $2p$ character. This is just one example illustrating 
that to date there is no consensus in the scientific community about the 
basic transport properties of doped manganites. It may be inferred, therefore,
that the understanding of these properties is far from completion and that 
further experimental and theoretical investigations of this matter are necessary. 
\par
It is well known that doped manganites of the same chemical composition but 
with different degrees of crystal lattice disorder show quite different  
transport and magnetic properties. The disorder can be altered either with 
variation of sample preparation conditions (for example, substrate temperature
and post-annealing at film preparation) or using radiation 
damage\cite{chen,wilson,stroud,ogale}. With increasing disorder the
resistivity peak temperature $T_{p}$ and the Curie temperature $T_{c}$ 
decrease, while the magnetoresistance increases. In understanding the 
nature of CMR  the influence of disorder of the crystal lattice is one 
of the important points and should be taken into account together with spin,
lattice and other effects. This communication is concerned mainly with this problem. 
\par
The object of investigation was thin-films  La$_{1-x}$Ca$_{x}$MnO$_{3}$ 
with $x \approx 1/3$. The disorder was enhanced by irradiating the films 
at room temperature with high-energy ($\simeq 6$~MeV) electrons. This high 
energy of the incident electrons makes it possible to produce a uniform 
distribution of damage defects, without any significant variation of defect
concentration as a function of depth (all incident electrons go through 
the film). In contrast to low energy ion irradiation, no interstitial implanted impurity
ions can remain in the film for electron irradiation to produce inhomogeneity. 
Similarly, in contrast to very high energy ion irradiation, electron irradiation
in our study does not produce extended defects, such as cascades and clusters. 
This facilitates the interpretation of the experimental results.  
At the low  damage level in this experiment, however, the electron radiation 
damage may indeed be quite similar to damage induced by very low level,
intermediately high-energy ion irradiation.   
\par
The maximal electron fluence in this study was about 
$2\times 10^{17}$~cm$^{-2}$.  The calculated quantity of displacements per 
atom (dpa) is about $10^{-5}$. This comparatively small radiation damage  
results in an appreciable increase in film resistivity. It was found that 
the relative resistivity increase in the ferromagnetic metallic state (below Curie 
temperature $T_{c}$) was much greater than in the insulating state (above $T_{c}$).
Such a small amount of radiation damage should not induce any noticeable 
resistance variations in ordinary ferromagnetic or non-ferromagnetic metals.
At the same time
any large influence of electron irradiation with the above-mentioned fluence 
on the $T_{c}$ and the magnitude of the colossal magnetoresistance was not observed.
Possible reasons for this unusual behavior for the doped manganites  are discussed.   

\section{Experiment and results}
\par
The La$_{1-x}$Ca$_x$MnO$_3$ films were prepared by physical
vapor codeposition of La, Ca and Mn from three separate,
independently controlled sources, similar to the technique for
preparation of Ca-Ba-Cu oxide precursors for growth of
oriented Tl$_2$Ca$_2$Ba$_2$Cu$_3$O thin films\cite{wang}.
The deposition was performed in 10$^{-5}$ Torr of oxygen onto 
LaAlO$_3$	substrates heated to
about 600$^{\circ}$C. La and Mn were evaporated from alumina
crucibles heated with a tungsten filament, and Ca was 
evaporated from a Knudsen cell. Post deposition anneals of the
films at 900$^{\circ}$C in flowing oxygen improved the CMR
behavior and produced well ordered films. The composition of 
the film was determined by microprobe analysis of an 
unannealed film deposited simultaneously onto a fused quartz
substrate. The films were also characterized by X-ray diffraction
and AC susceptibility measurements. Agreement among the values of 
$T_{c}$ determined by the real part $\chi^{'}$ of the susceptibility
and $T_{p}$ determined by both measurements of the resistivity and the 
imaginary part $\chi^{''}$ of the susceptibility confirm that the films 
have good chemical and magnetic homogeneity based on the scheme proposed
by Araujo-Moreira, $\it et~al.$\cite{ara}. Further details of the
preparation technique and characterization are presented
elsewhere\cite{don}.
\par
Although a sensitive magnitometer was not available for magnetization
measurements with these films, AC susceptibility was measured, both for
unirradiated and irradiated films. In each case the onset of the sharp increase
in the real part of the susceptibility $\chi^{'}$ and the sharp peak in the
imaginary part of the susceptibility $\chi^{''}$ coincide within experimental
error with the value of $T_p$. The sharp increase in the low frequency
$\chi^{'}$ ($\approx 140$~Hz) data presumably corresponds to the magnetic
transition temperature $T_c$. Representative data for an unirradiated film is
presented in Fig.~4 of Ref.\ \onlinecite{don}. Data for $\chi^{'}$ and
$\chi^{''}$ for one of the films irradiated in this study (not shown) has much
less noise and provides clear evidence that $T_c$ and $T_p$ coincide, both for
the unirradiated and irradiated films in this study. This is not unexpected,
however, since ion irradiation studies have shown\cite{stroud} that for high
quality films with small lattice damage, these two temperatures are essentially the
same, but  for much higher lattice damage $T_p$ will be at much lower
temperature than $T_c$. Throughout this paper reference will be made to $T_p$,
but, since $T_p$ and $T_c$ are essentially identical, the conclusion from these
experiments apply to both equally well. 
\par
During the electron irradiation the films were in a special holder which
was cooled with running water and a powerful fan. The estimated 
overheating above room temperature during the irradiation was no more 
than $\simeq 15^{\circ}$~C. Two film samples were investigated 
(x=0.27 and 0.36). These films (with thicknesses about 300 nm) were 
prepared under nearly the same conditions.
The resistance of the films was measured using a standard four-probe 
technique. An applied magnetic field (up to 20 kOe) was perpendicular to 
the film plane and to the direction of current. The results obtained 
were nearly the same for both films and will be illustrated by the 
data from the x=0.36 sample.  The transport properties of 
this film in its initial state (before irradiation) correspond to the 
usual behavior of CMR films (Fig.1 and 2). Namely, the temperature 
dependence of resistance $R(T)$ has a maximum (peak) at $T_{p} \approx 280$~K 
(the maximum is rather smeared). Below $T_{p}$ (which for these manganite
samples is always in the vicinity of the Curie point $T_{c}$) the 
temperature behavior of the resistance is metallic in character. The 
resistance $R_{p}$ at $T_{p}$~is about 1315 $\Omega$ (this corresponds
to the resistivity $\rho = 1.24\times 10^{-2}$~$\Omega$cm); 
whereas, already at $T = 200$~K the resistance $R_{200}$ is much less 
(178~$\Omega$), and at $T = 120~$K the resistance has decreased to 
$R_{120} \approx 66 $~$\Omega$
($\rho = 6.25\times 10^{- 4}$~$\Omega$cm). 
We have taken $\delta_{H} = [R(0) - R(H)]/R(H)$ 
at a magnetic field $H = 16$~kOe as a measure of the magnetoresistance. 
It can be seen from Fig.3 that $\delta_{H}$ has its maximum value (about 
66~\%) at a characteristic temperature $T_{m} \approx 265$~K 
($T_{m}$ is also near $T_{c}$ for these manganites). 
\par
After the first irradiation with a fluence $\Phi \approx 9\times 
10^{16}$~cm$^{-2}$ the above mentioned parameters have changed 
to the following values:
$T_{p} \approx 278$~K, $R_{p} \approx 1480$~$\Omega$, 
$T_{m} \approx 259$~K, $\delta_{H} = 65$~\%,  
$R_{200} = 266 \ \Omega$, $R_{120} = 130 \ \Omega$ (Figs. 2 and 3).
After a second irradiation (the total fluence after two irradiations 
is about $2\times 10^{17}$~cm$^{-2}$) these parameters are: 
$T_{p} \approx 275$~K, $R_{p} = 1670$~$\Omega$,
$T_{m} \approx 261$~K, $\delta(H) = 64$~\%, $R_{200} = 323 \ \Omega$, 
$R_{120} = 191 \ \Omega$.                    
\par
It can be seen from these results that the electron irradiation 
has produced a rather large effect on film resistance. The film 
resistance in the paramagnetic insulating state (above $T_{p}$) has 
increased over 25 \%. More striking is the change in $R$ in the
ferromagnetic state at low temperature: $R(120)$ is tripled by the 
electron irradiation. At the same time (taking into account 
the experimental errors) there is no substantial changes of the CMR characteristics: 
the values of $T_{p}$, $T_{m}$ (and thus $T_{c}$) decrease only about 
5~K; whereas, the magnitude of the magnetoresistance $\delta_{H}$ remains
practically unchanged.  
\par
In discussion and analysis of the results obtained it is important to 
determine the  degree of radiation damage produced by the electron 
irradiation in our study. The types of defects produced by electron irradiation
are Frenkel pairs, i.e. isolated vacancies and interstitials. The atomic 
displacement cross sections by 
fast electrons and the corresponding values 
of dpa for all elements (La, Ca, Mn, and O) of the sample have been 
calculated taking into account the exact chemical composition of the 
film and using the well-known fundamental concepts of such type of 
relativistic calculations \cite{seitz,dien} and the cascade calculational
procedures outlined in Ref.\ \onlinecite{oen} together with the ratios of
the Mott to the Rutherford cross section $M(x,E)$. The results of this type
of calculation depend essentially on the specified value of the threshold 
energy $E_{d}$ (an atom which receives energy $E \geq E_{d}$ will be 
displaced certainly from its lattice site\cite{seitz,dien}) which was
chosen to be $E_{d} = 20$~eV for all ions, the 
typical value of $E_{d}$ in common use for this type of calculation.  
\par
At the total fluence $\Phi \approx 2\times 10^{17}$~cm$^{-2}$ the 
calculations result in the following values of dpa for the chemical
elements which comprise this film: 
$3.2\times 10^{-5}$~(La), $2.2\times 10^{-6}$~(Ca), $9.3\times 10^{-6}$~(Mn),
$3.4\times 10^{-6}$~(O). The total dpa is about $4.7\times 10^{-5}$. 
One should not take these values literally.
As mentioned above, the output of such calculations depends essentially
on the values of energy $E_{d}$, which are obscure and which may be quite
different for the different constituent elements. Nevertheless, we believe,
based on previous 
studies\cite{seitz,dien,oen}, the calculation results should be correct at 
least to the order of the magnitude.

\section{Discussion}
The experimental results correlate, at least qualitatively,  with  the DE 
model \cite{zener,and55,gennes}. In 
this model the ferromagnetic coupling between pairs of Mn$^{3+}$ and 
Mn$^{4+}$ ions through the oxygen ions is also responsible for the 
metallic properties of the manganites. The electron hopping amplitude 
$t_{i,j}$ from site $i$ to site $j$ is given by
\begin{equation}
t_{i,j} = b_{i,j}\cos (\theta_{i,j}/2), 
\label{1}
\end{equation}
where $b_{i,j}$ is a material-dependent constant, $\theta_{i,j}$ is the angle
between the directions of two ionic spins. It can be seen from Eq.\ \ref{1} 
that in the DE model a clear connection exists between electron transport and
magnetic order, i.e. the electron conduction is a function of magnetic order. 
The angle $\theta_{i,j}$ decreases below $T_{c}$ or in a magnetic field. This
may be a possible reason for CMR. The disorder (for example,
vacancies) must reduce the coupling between the 
Mn$^{3+}$--O--Mn$^{4+}$ ions and, therefore, the probability of electron 
transfer. This must cause the increase in resistivity. At the same time 
the disorder should influence the ferromagnetic order (the Curie temperature
$T_{c}$ must go down). Therefore, the increase of disorder must induce 
simultaneously an increase of resistance and decrease of $T_{c}$ in the 
DE model, that qualitatively corresponds to our results and the results 
of previous studies with ion-irradiated manganites\cite{chen,wilson,stroud}.
\par
It is usually assumed that $b_{i,j}$ in Eq.\ \ref{1} is a constant for all 
lattice cells, which can be correct only in perfect crystals. It was taken 
into account in Ref.\ \onlinecite{stroud} that in disordered crystals 
$b_{i,j}$ is a position-dependent quantity and denotes a static disorder.
The numerical simulations in  Ref.\ \onlinecite{stroud} in the frame of 
the model for disorder-induced polaron formation\cite{emin} have shown 
that increasing static disorder decreases the values of $t_{i,j}$ and 
leads to a metal-insulator transition as observed in 
Refs.\ \onlinecite{chen,wilson,stroud,ogale}. 
\par
The general approach of Ref.\ \onlinecite{stroud} (to look beyond the 
DE model and take into consideration additional important effects)
seems to be  quite fruitful. 
The proper consideration and interpretation of the irradiation-disorder 
influence is possible, however, only if the exact conduction mechanisms 
in the insulating and high-conducting ferromagnetic regimes of the doped 
manganites are known. At the moment there is still no clear enough 
understanding of these mechanisms. Nevertheless the experimental and theoretical
achievements in this matter in the last years\cite{nagaev,coey,ram,khom,millis,gor}
enable such an attempt. 
\par 
Some general observations should be noted. The magnitude  of the resistance
increase near and above $T_{p}$  
(about 25 \%) at first sight does not arouse great surprise, since semiconductors
with a very small concentration of charge carriers are generally very sensitive
to irradiation that produces displacement atoms in the crystal. The irradiation 
defects quite often cause the reduction of  charge carrier concentration and 
mobility\cite{seitz,dien,james}. The
charge carrier concentration in doped manganites is not, however, very small.
Based on the chemical doping the charge carrier concentration in  
La$_{1-x}$Ca$_{x}$MnO$_{3}$~($x \approx 0.33$) should be about 0.33 holes per
unit cell, a density of carriers $n \approx 6\times 10^{21}$~cm$^{-3}$ 
(for the cubic cell with lattice parameter about 0.385 nm). In Hall-effect 
studies of this compound it was found that in ferromagnetic state below
$T_c$ the  charge-carrier density should be in the range 0.85-1.9 per unit
cell\cite{snyder,jacob}. Even higher value (2.4) was found in Ref.\ 
\onlinecite{chun} for La$_{2/3}$(Ca,Pb)$_{1/3}$MnO$_3$. The reasons for such
high values (which deviate much from the nominal doping level) is not clear at
present\cite{jacob,chun}. Because of this we will assume that charge-carrier
density in the ferromagnetic state corresponds roughly to 0.33 holes per unit 
cell ($n \approx 6\times 10^{21}$~cm$^{-3}$). In the paramagnetic state not all     
the dopants contribute to the charge carrier density. Part of the doped holes
may be localized\cite{nagaev,gor}. Indeed, it follows from the Hall-effect 
measurements above $T_c$ that in the  paramagnetic insulating state the 
charge-carrier density is much lower, namely, in the range from 0.004 to 0.5 
holes per unit cell\cite{jacob,chun,nunez,li}. We can rather safely assume 
that charge-carrier density below $T_c$ decreases by at least a factor of 
five. This corresponds approximately to the value 
$n \simeq 10^{21}$~cm$^{-3}$ which can be used for numerical evaluations. In 
the case of a semiconductor 
with activated conductivity due to a band gap or mobility edge this value 
appears to be too high to understand how the $10^{-5}$ dpa can produce this
rather appreciable (about 25 \%) resistivity
increase. Indeed, it is easy to see that even if each of the displaced ions 
produces a trap for the mobile charge carrier, the traps can lead to localization of
only about $4\times 10^{18}$ carrier/cm$^{3}$ which is much less than estimated 
carrier density. Therefore, the explanation based on the reduction in charge 
carrier density,  which is quite usual for  semiconductors\cite{seitz,dien,james},
cannot explain the observed irradiation induced resistance increase for these manganites. 
\par
It is even more difficult to explain how such a low dpa can induce the observed 
threefold increase in the resistivity in the metallic ferromagnetic
state at low temperature (Fig. 2). It is known\cite{seitz,dien,james},
that 1\% of displaced atoms (that is 0.01 dpa) in the noble metals like Au,
Ag or Cu result in a change of resistivity of about 1~$\mu \Omega$~cm.  Such
small changes practically could not be experimentally distinguished for these 
rather high-resistance manganite films. It follows  that additional assumptions
which take into account the peculiarities of the 
insulating and metallic states and the nature of the charge carriers in doped
manganites are needed to explain the experimental results. For the insulating
state of the doped manganites it is essential to take into account the polaronic
nature of charge carriers in them 
(see Refs.\ \onlinecite{ram,millis,gor,alex,roder,ibar,hund,polar,ziese}
and references therein). The introduction to polaron physics and the main 
references can be found, for example, in Ref.\ \onlinecite{mott}. It is rather
commonly assumed that the conductivity of doped manganites
above $T_{c}$ is determined by small polarons\cite{hund,polar,ziese}. 
The exact nature of these small polarons is now the object of intensive
theoretical and experimental studies. The different kinds of  lattice or
magnetic polarons are considered. It is widely accepted that at decreasing 
temperature in the region of $T_{c}$ the crossover from
localized small polarons to itinerant large polarons takes place\cite{boris}. 
This point of view has found experimental
support\cite{hund,lanz}.
\par
According to the definition\cite{mott}, a lattice polaron is the unit
consisting of the ``self-trapped'' (localized) charge carrier, together 
with its induced lattice deformation. The polaron is called small when the 
spatial extent of the wave function of the trapped charge carrier is
comparable with the separation of next-neighbor ions. The polaron radius
$r_{p}$ for small polarons in doped manganites is estimated to be about 
0.5 nm\cite{ziese}. Small polarons have a large effective mass (10-100 
larger than mass of free electron) and can move by tunneling or thermally
activated hopping. The mobility of the small polaron is very low because 
the charge carrier movement includes the displacements of atoms surrounding it.
\par
For any conductor the conductivity $\sigma$ is given by
the general relation $\sigma = ne\mu$, where $n$ is density of carriers
and $\mu$ is mobility. In contrast to band semiconductors in which $n$ can
depend on temperature in a thermally activated way, the density of
carriers is assumed to be constant with temperature for polaronic
conductors. At fairly high temperatures $T > \theta_{D}/2$ (where $\theta_{D}$ 
is the Debye temperature) in the adiabatic limit\cite{mott} (which is assumed 
to be true for the doped manganites\cite{ziese,wor})
it is the small polaron mobility that is activated and the 
resistivity $\rho = 1/\sigma$ is given by
\begin{equation}
\rho = \frac{2kT}{3ne^{2}a_{h}^{2}\omega_{0}}\exp(E_{a}/kT), 
\label{2}
\end{equation}
where $E_{a}=E_{b}/2 - J$ is the activation energy, with $E_{b}$ the
polaron binding energy and $J$ the overlap integral; $a_{h}$ is the hopping
distance, and $\omega_{0}$ is the optical-phonon frequency. 
\par 
Eq.\ \ref{2} is true in the dilute, noninteracting limit, when the density 
of carriers is far less than the density of equivalent hopping sites\cite{mott,wor}.
It may be assumed as in Ref.\ \onlinecite{ziese}, that in doped
manganites all the carriers form polarons. In this case with the 
above-estimated value of charge carrier density in the insulating
paramagnetic state ($n \simeq 10^{21}$~cm$^{-3}$) the mean distance 
$l_{ch}$ between the trapped charge carriers is $\approx 1.0$~nm. Since 
it is assumed\cite{mott} in the general case that the hopping distance $a_{h}$ 
is equal to a lattice constant, the noninteracting limit is quite justified for
these doped manganites.  For the value of dpa in this  study
(about $5\times 10^{-5}$) the mean distance $l_{d}$ between the damage
lattice sites is about 6 nm. In Ref.\ \onlinecite{stroud} a 
much larger dpa (about 0.01) was produced by ion irradiation.
This resulted in a tenfold increase in the resistivity in the insulating state, 
as compared to the approximately $25\%$ increase shown in Fig.~2 for 
electron irradiation. In that experiment the length $l_{d}$ would be 
approximately 1.0 nm. 
\par
The effect of radiation damage in the insulating state of doped manganites 
can be understood, at least qualitatively, by  taking into account the 
small-polaronic nature of charge carriers. Two main sources of radiation 
influence on small polaron conduction in doped manganites can be seen. First, 
according to Ref.\ \onlinecite{emin}, for the crystals with not too strong 
an electron-lattice interaction it is quite possible that some appreciable 
number of carriers would be quasifree rather than small polarons. This should
be true for the doped manganites since  many experimental and theoretical 
studies indicate\cite{polar,lanz,booth,louca,igor} the coexistence of 
localized and itinerant carriers   in a rather wide temperature range near $T_{c}$.
In this case the disorder can convert some of the available quasifree states to
small-polaron states\cite{emin}.  That is, disorder reduces the strength of the
electron-lattice coupling needed to stabilize the global small-polaron
formation. Defects and impurities serve as centers for electron
localization and small-polaron formation. 
This explanation is supported  by the numerical simulations in 
Ref.\ \onlinecite{stroud}. This mechanism of the disorder-induced conductivity 
decrease may be dominant near $T_{c}$. 
\par
In ion-irradiation experiments\cite{chen,wilson,stroud,ogale} much larger dpa values
(up to 0.01 and more) have been produced which have resulted in an increase in
resistance in the insulating paramagnetic state by one (and sometimes two) order
of magnitude. This effect is accompanied by an increase in the activation energy 
$E_{a}$ (see Eq.\ \ref{2}) and a decrease in peak temperature $T_{p}$. In this 
case, especially at temperatures rather far above $T_{c}$, it is not possible to
explain the resistance increase only by the transformation of available quasi-free
carriers to small polarons. 
These results demonstrate that the disorder influences directly the charge-carrier 
hopping and leads to a decrease in the charge-transfer probability.  There
appear to be  no specific theoretical treatments of this problem for small-polaron
hopping.  It is known that at high temperatures polaron jumps occur when electron
energies associated with the initial and final sites (these energies are determined
by a configuration of lattice atoms) are equal\cite{mott}. Maybe disorder affects
these so called coincidence events in such way that it leads to a decrease in 
transfer probability. It should be taken into account also the possible influence
of Anderson localization\cite{nagaev,coey}. It is evident that more experimental
and theoretical efforts are needed to clarify this problem.   
\par
The foregoing discussion indicates that an adequate 
consideration of radiation-damage effects on conductivity is possible
only in the frame of a rather strictly determined conduction mechanism
and charge-carrier nature. Unfortunately, no determination has  been made for the
ferromagnetic high-conducting state of doped manganites well below $T_{c}$.
At least one assumption for this state is, however, clear: the charge carriers
at low temperatures can be considered to be quasifree. It has been 
argued\cite{millis,roder,hund,lanz} that the charge carriers
in this state are itinerant large polarons. The polaron of this 
type\cite{mott} moves without thermal activation and behaves like a heavy 
particle (with mass in 2-4 times larger than mass of free electron). Another
possibility is that the doped manganites below $T_{c}$ are just degenerate
semiconductors\cite{nagaev}. In any case the doped manganites in the
ferromagnetic state with a minimal resistivity of about 100~$\mu \Omega$~cm  
should be considered as some kind of
``bad'' metal, like heavily doped semiconductors or amorphous metals. For 
such conductors it is quite  difficult (and sometimes of no use) to estimate
a  value of the electron mean-free path $l$ and consider the decrease of $l$
under influence of irradiation-induced disorder. Indeed, for a Fermi 
velocity $v_{F} = 7.6\times 10^{5}$~m/s (as was calculated in 
Ref.\ \onlinecite{pick} for La$_{0.67}$Ca$_{0.33}$MnO$_{3}$) the use 
of the quasifree-electron relation $1/\rho = ne^2\tau/m$ with $n$ about 1.0
hole per unit cell and $m$ given by the mass of a free electron, gives
$l = v_{F}\tau \approx 0.25$~nm. With an effective mass $m^{*} = 4m$,
$l \approx 1$~nm. The films in this experiment are not single-crystal,
but they do consist of rather large grains with a size near 0.5 $\mu$m.
Therefore, the ``intrinsic'' value of $l$ within the grains determined 
in this model  should be larger. It is inconceiveable, however, that such 
considerations  with a mean distance $l_{d}$ between the damage sites of
about 6 nm could explain  the threefold increase in the resistivity of such
a rather ``bad'' metal with an electron mean-free path on the order of 1 nm.
\par
The unusual magnetic behavior of the doped manganites suggests a possible
phenomenological explanation of the large effect of small radiation damage on
the resistance in the ferromagnetic metallic regime. Irradiation not only leads
to lattice disorder that can lead to elastic electron scattering as in normal
non-ferromagnetic metals, but it also perturbes the long-range ferromagnetic
order. In the manganites the conductivity increases with the enhancement of
ferromagnetic order. Indeed, that is the source of the huge resistivity decrease
at the paramagnetic-ferromagnetic transition and the CMR. Below $T_c$ an
unusual correlation between resistivity and magnetization $M(T,H)$ has been
reported \cite{hund,hef}.
For example, in Ref.\ \onlinecite{hund} the following experimental relation
between $\rho$ and $M(T,H)$ for the  La$_{0.7}$Ca$_{0.3}$MnO$_{3}$ films was found
\begin{equation}
\rho (T,H) = \rho_{m} \exp \{{-M(T,H)/M_{0}}\}, 
\label{3}
\end{equation}
where  $\rho_{m}$ and $M_{0}$ are  sample-dependent parameters. At present 
there is no clear theoretical understanding of this correlation between $\rho$ 
and $M$. It is generally accepted that the increase in $M$ should lead to 
delocalization of the holes and to the increase in hole mobility. In any case,
however, it is clear that doped manganites are not conventional
ferromagnetic metals even well below $T_{c}$, and that electronic
transport in them is influenced to a high degree by magnetic
order\cite{hund}.    
\par
A reasonable hypothesis is that the dominant effect of irradiation on the
resistivity of the doped manganites at low temperature in the ferromagnetic
phase comes primarily from the disruption of long range magnetic order, perhaps
through the magnetoelastic coupling that produces magnetostriction. Indirect
evidence for this is provided by the observation that ion irradiation induces a considerable decrease in the 
saturation magnetization value $M_{s}$\cite{stroud,ogale}. For example,
in Ref.\ \onlinecite{ogale} for an ion irradiation dose which has resulted
in the nearly same dpa ($\simeq 10^{-5}$) as in the present study, the 
saturation magnetization decreased by about $30\%$.
Further indirect evidence of the influence of disorder effects on $M_{s}$ is
provided by the three-fold decrease in $M_{s}$ with only a small shift in $T_c$
that was associated with a decrease in grain size from 110 to 20 nm in bulk
samples\cite{sanchez}. Unfortunately, the additional experimental facilities
needed to test this hypothesis were not available for this experiment, but its
discussion may lead to future tests of the hypothesis and generate new interest
in irradiation damage studies as a way to probe the fundamental nature of
conduction in these exotic materials.
\par
In conclusion, the high-energy electron irradiation effect on the transport
properties of La$_{1-x}$Ca$_{x}$MnO$_{3}$ films ($x \approx 1/3$) has been
investigated. Comparatively small electron fluences used in this  study do
not have any substantial influence on the Curie temperature $T_{c}$ or the
magnitude of the magnetoresistance. At the same time these fluences result
in an appreciable increase in film resistivity in both the insulating 
paramagnetic state and especially in the highly conductive ferromagnetic state.
The relative resistivity increase in the metallic ferromagnetic state 
(below $T_{c}$) was found to be much (an order of magnitude) greater than that
in the insulating paramagnetic state. This behavior is quite different from
that associated with  non-magnetic metals and semiconductors and  can be
understood in the high-temperature regime qualitatively by taking into account 
the polaronic nature of manganite's conductivity above and near $T_{c}$. A
possible explanation for the low temperature behavior has been suggested, but it
must be tested with magnetization measurements that were not available to the
present experiments. 
\acknowledgments
The work at Texas A\&M is supported by the Robert A. Welch Foundation (A-0514),
the Texas Advanced Technology Program (Grant No. 010366-141  ), and the 
Texas Center for Superconductivity at the University of Houston (TCSUH). BIB
and DGN acknowledge support by NATO Scientific Division (Collaborative Research
Grant No. 972112). The authors wish to thank I. Lyuksyutov, V. Pokrovsky and 
E. L. Nagaev for useful discussions and C. W. Chu at TCSUH for X-ray diffraction
of the samples. The authors are especially grateful to S. M. Seltzer of the 
National Institute of Standards and Technology for the calculation of the 
Mott/Rutherford ratio $M(x,E)$ for the La-Ca-Mn-O compounds and to R. M. Stroud
for copies of her work before publications.

\begin{figure}
\caption{Temperature dependence of the resistivity of a 
non-irradiated La$_{0.64}$Ca$_{0.36}$MnO$_{3}$ film on a LaAlO$_{3}$
substrate for different magnitudes of applied magnetic field.}
\label{Fig.1}
\end{figure}

\begin{figure}
\caption{Temperature dependence of the resistivity of a 
La$_{0.64}$Ca$_{0.36}$MnO$_{3}$ film on a LaAlO$_{3}$ substrate 
for different degrees of radiation damage: $\circ$ --- initial state, 
${\scriptstyle \triangle}$ --- after irradiation with
6MeV electrons at a fluence 
$\Phi \approx 9\times 10^{16}$~cm$^{-2}$, $\bullet$ --- after 
irradiation at a total fluence $\Phi \approx 2\times 10^{17}$~cm$^{-2}$.} 
\label{Fig.2}
\end{figure}

\begin{figure}
\caption{Temperature dependence of the magnetoresistance $\delta_{H}$ 
for the La$_{0.64}$Ca$_{0.36}$MnO$_{3}$ film on a LaAlO$_{3}$ substrate 
for different degrees of radiation damage: Symbols are the
same as in Fig.~2.}
\label{Fig.3}
\end{figure}


\begin{references}
\bibitem{boris}In our discussion we will assume that small or large polarons,
as well as lattice or magnetic polarons are charge carriers according to the
definitions in Ref. \ \onlinecite{mott}. At the present time in some scientific
papers the term ``large polaron'' is sometimes used as a name for ferromagnetic
droplets or clusters resulting from phase-separation or inhomogeneity effects
in the doped manganites.
\end{references}
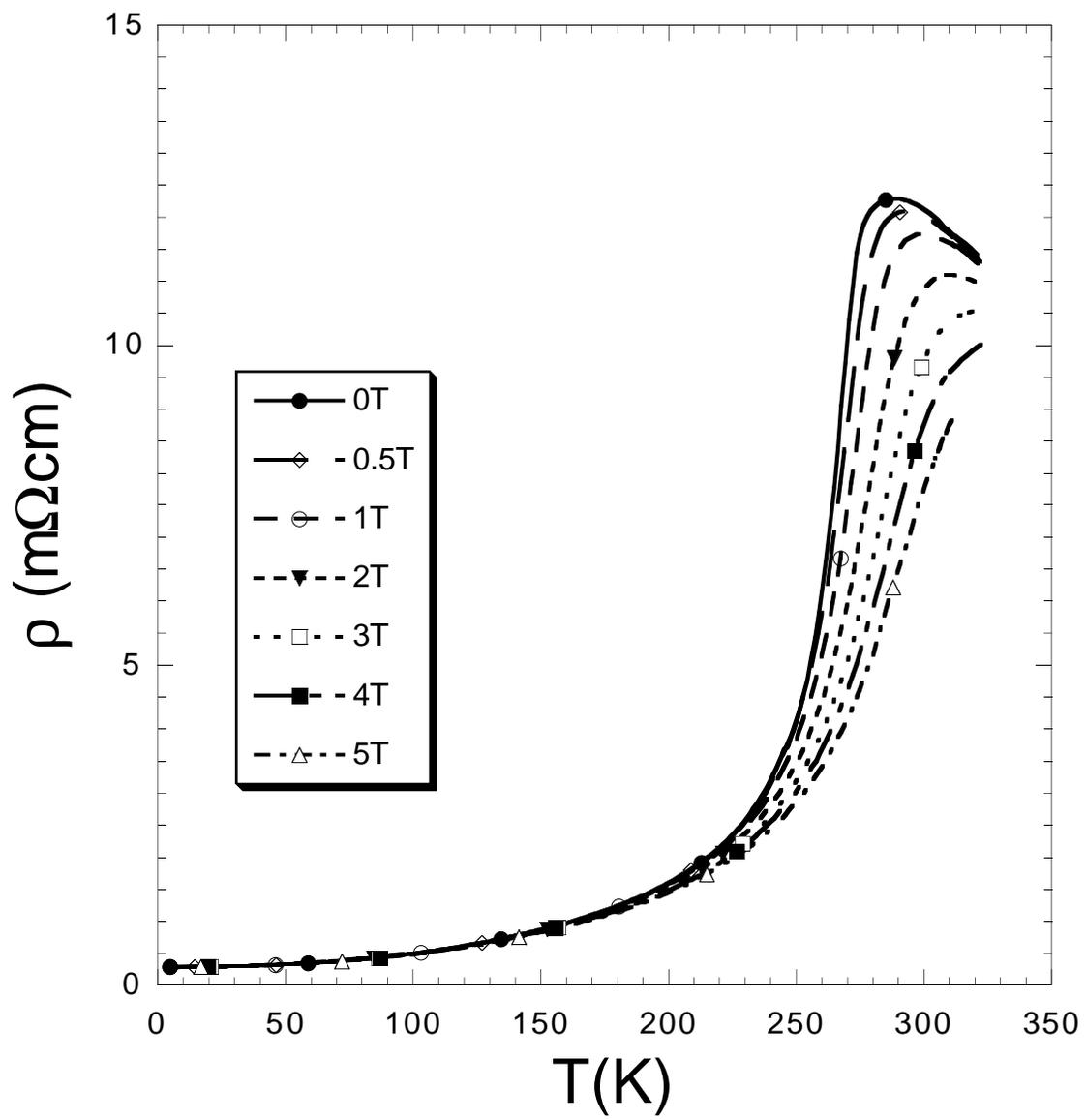
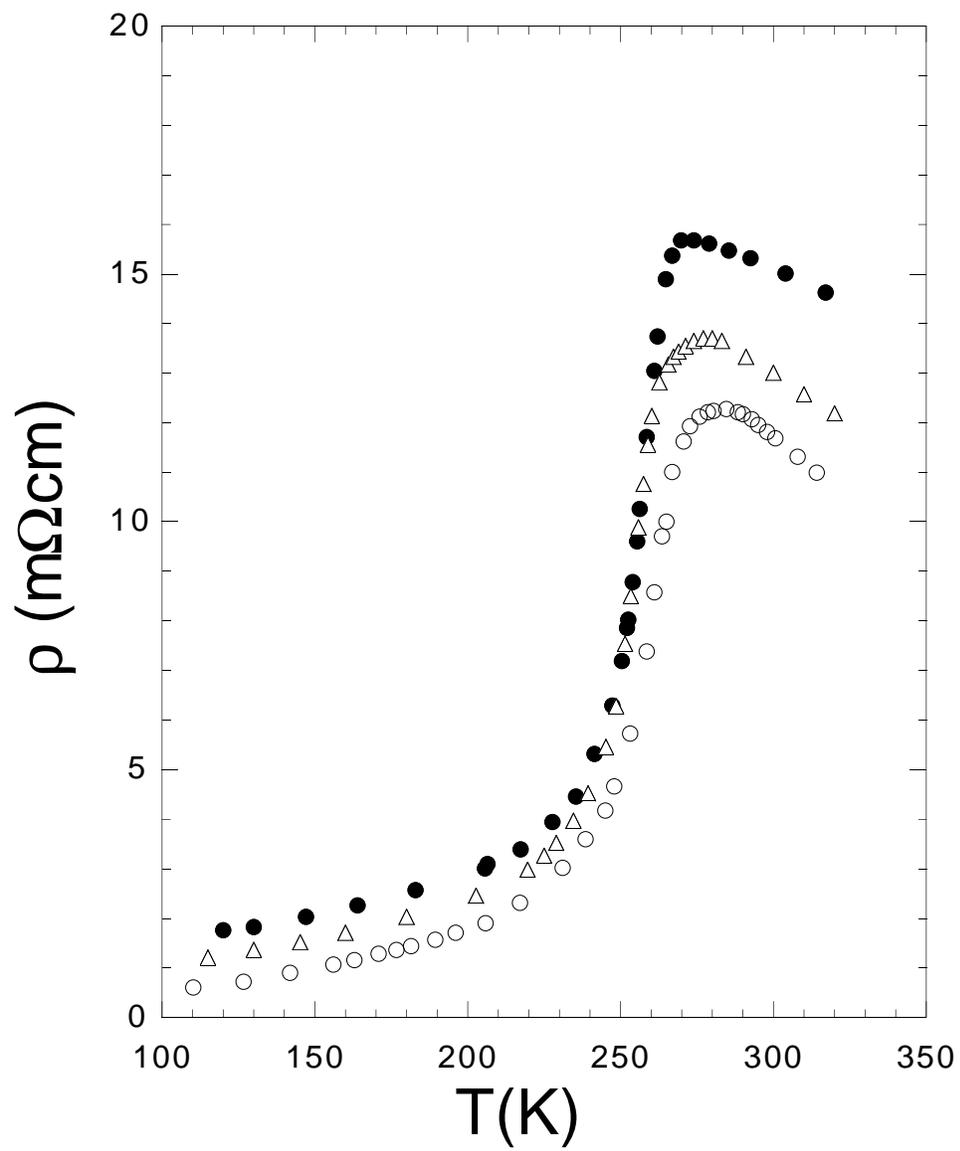
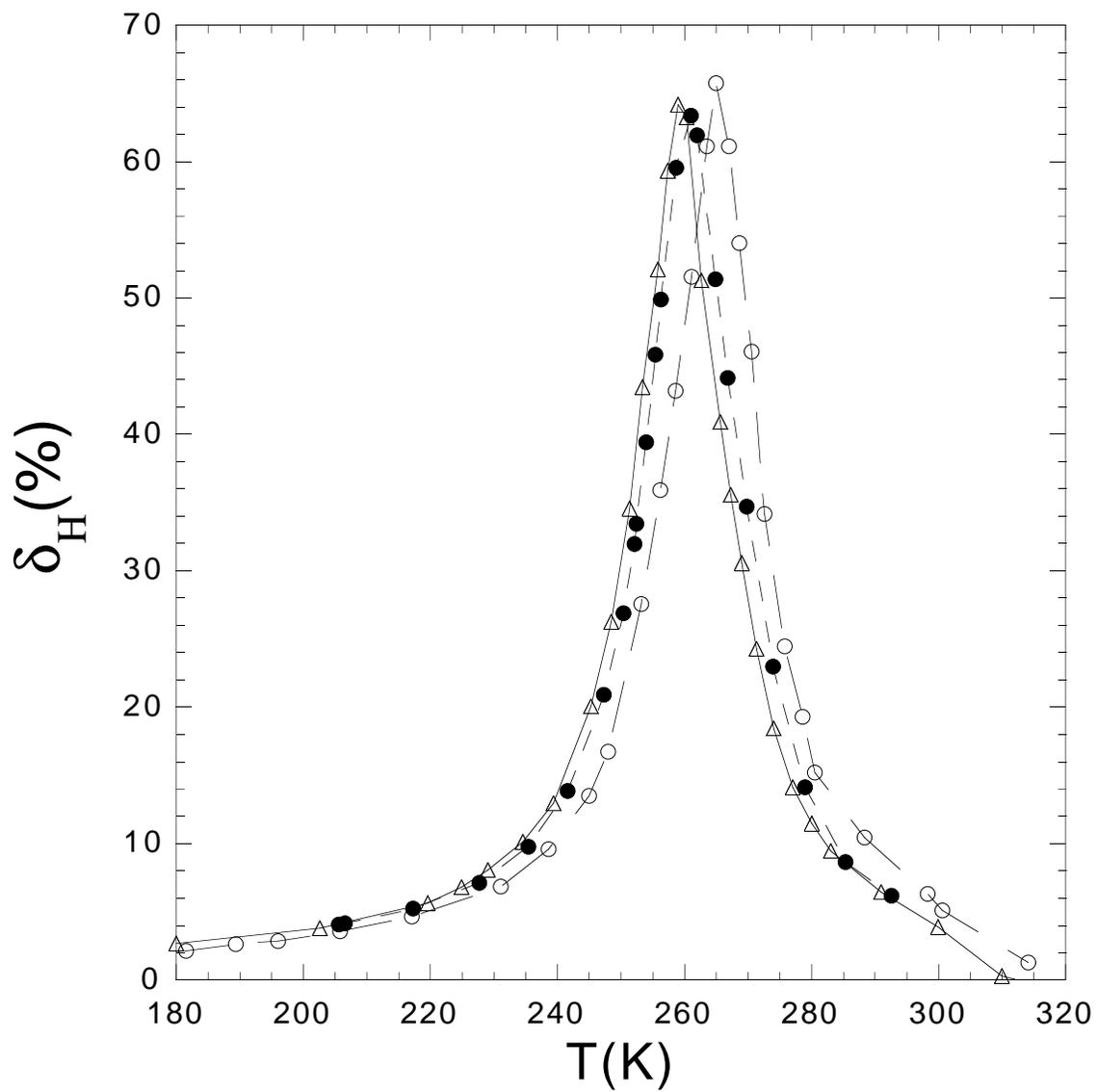
\end{document}